\documentclass[pdflatex,sn-mathphys-num]{sn-jnl}

\usepackage{graphicx}%
\usepackage{multirow}%
\usepackage{amsmath,amssymb,amsfonts}%
\usepackage{amsthm}%
\usepackage{amssymb}
\usepackage{mathrsfs}%
\usepackage[title]{appendix}%
\usepackage{pifont}
\usepackage{caption}
\usepackage{xcolor}%
\usepackage{indentfirst}
\usepackage{textcomp}%
\usepackage{longtable}
\usepackage{manyfoot}%
\usepackage{booktabs}%
\usepackage{algorithm}%
\usepackage{hyperref}
\usepackage{lineno}  
\usepackage{algorithmicx}%
\usepackage{algpseudocode}%
\usepackage{listings}%
\newcommand{\arcsec}{\mbox{\ensuremath{^{\prime\prime}}}}

\theoremstyle{thmstyleone}%
\theoremstyle{thmstyletwo}%

\theoremstyle{thmstylethree}%

\begin{document}

\title[An X-ray-Emitting Proto-Cluster at $z\approx5.7$]{An X-ray-Emitting Proto-Cluster at $z\approx5.7$ Reveals Rapid Structure Growth}


\author*[1]{\fnm{\'Akos} \sur{Bogd\'an}}\email{abogdan@cfa.harvard.edu}

\author[1]{\fnm{Gerrit} \sur{Schellenberger}}

\author[2]{\fnm{Qiong} \sur{Li}}

\author[2]{\fnm{Christopher J.} \sur{Conselice}}

\affil*[1]{Center for Astrophysics \ding{120} Harvard \& Smithsonian, 60 Garden Street, Cambridge, MA 02138, USA}

\affil[2]{Jodrell Bank Centre for Astrophysics, University of Manchester, Oxford Road, Manchester UK}

\abstract{
\mathversion{normal}
Galaxy clusters are the most massive gravitationally bound structures in the universe and serve as tracers of the assembly of large-scale structure  \cite{2012ARA&A..50..353K}. Studying their progenitors, proto-clusters, sheds light on the earliest stages of cluster formation. Yet, detecting proto-clusters is demanding: their member galaxies are loosely bound and the emerging hot intracluster medium (ICM) may only be in the initial stages of virialization \cite{2013ApJ...779..127C,2016A&ARv..24...14O,2017ApJ...844L..23C}. Recent \textit{JWST} observations located several proto-cluster candidates by identifying overdensities of $z\gtrsim5$ galaxies  \cite{2022A&A...667L...3L,2023ApJ...947L..24M,2024ApJ...974...41H,2025MNRAS.539.1796L,2025MNRAS.tmp...36H}. However, none of these candidates was detected by X-ray observations, which offer a powerful way to unveil the hot ICM. Here, we report the combined \textit{Chandra} and \textit{JWST} detection of a proto-cluster, JADES-ID1, at $z\approx5.68$, merely one billion years after the Big Bang. We measure a bolometric X-ray luminosity of $L_{\rm bol} = (1.5^{+0.5}_{-0.6}) \times10^{44} \ \rm{erg \ s^{-1}}$ and infer a total gravitating mass of $M_{500}= (1.8^{+0.6}_{-0.7}) \times 10^{13} \ \rm{M_{\odot}}$, making this system a progenitor of today's most massive galaxy clusters. The detection of extended, shock-heated gas indicates that substantial ICM heating can occur in massive halos as early as $z\approx5.7$. In addition, given the limited survey volume, the discovery of such a massive cluster is statistically unlikely \cite{2008ApJ...688..709T}, implying that the formation of the large-scale structure must have occurred more rapidly in some regions of the early universe than standard cosmological models predict.}


\maketitle

Proto-clusters are the earliest phase in the assembly of large-scale cosmic structures. These systems mark a transition from initial galaxy overdensities to fully virialized galaxy clusters. X-ray and Sunyaev-Zel'dovich (SZ) observations have been widely used to identify clusters and proto-clusters in the early universe, up to $z\sim2.5$, that are either gravitationally collapsed or in the process of collapsing \cite{2009ApJ...701...32S,2011A&A...526A.133G,2016ApJ...817..122B,2018A&A...620A...2M,2020ApJ...902..144D,2020MNRAS.496.1554M,2020ApJS..247...25B,2021ApJS..253....3H,2023Natur.615..809D}.  However, detecting proto-clusters within the first billion years of the universe ($z\gtrsim5.5$) remains a major challenge. At these early epochs, halos may not yet have experienced significant virial heating (i.e.\  shocks that bring the gas up to the virial temperature of the halo) or may only be in their earliest stages \cite{2012ARA&A..50..353K,2016A&ARv..24...14O}. This makes observing the faint hot ICM, the key signature of the onset of virial heating, difficult. Therefore, identifying the first proto-clusters undergoing virialization necessitates a multi-wavelength effort. While overdensity measurements reveal dense galaxy environments, unambiguous confirmation of the onset of gravitational collapse comes from the detection of ICM \cite{2011NJPh...13l5014F,2020A&A...642A.124T,2024A&A...691A.300X}. In this work, we extend this approach by utilizing \textit{JWST} and \textit{Chandra} data to probe an even earlier proto-cluster, merely one billion years after the Big Bang. 

\textit{JWST} observations have revolutionized the study of high-redshift structures by identifying substantial populations of faint and distant galaxies \cite{2023Natur.616..266L,2023NatAs...7..731B,2022ApJ...938L..15C}. Its near-infrared imaging capability and accurate photometric redshifts allow the identification of galaxy overdensities, paving the way for detecting the earliest proto-clusters. In the JADES field \cite{2023arXiv230602465E}, one of \textit{JWST}'s premiere regions, six proto-cluster candidates were identified in the redshift range of $z=5-7$ \cite{2025MNRAS.539.1796L}. Of these proto-clusters, JADES-ID1 is the most compelling candidate, exhibiting high galaxy overdensity and cluster membership likelihood. A total of 66 galaxies are identified as potential members, with an inferred halo mass of $\rm log(M_h/M_{\odot})$ = 13.28$^{+0.37}_{-0.34}$, the highest among the JADES fields \cite{2025MNRAS.539.1796L}. Because only JADES-ID1 has the richness and halo mass to produce a detectable ICM signal in the JADES field, our X-ray analysis targets this proto-cluster candidate. The CDFS/JADES field uniquely combines deep \textit{JWST} imaging with the deepest \textit{Chandra} exposure over a single footprint, making it currently the only survey volume in which X‑ray emission from the ICM of high‑redshift proto‑clusters could in principle be detected. Further details on the identification of JADES-ID-1 and quantitative tests ruling out the presence of projected foreground galaxy groups and clusters are presented in Methods Section 1.

Within a projected radius of $42\arcsec$ ($\approx250$~kpc) around the JADES-ID1 centroid, we measure the local galaxy overdensity of $\delta_{gal} = 3.9$ relative to the mean galaxy density across the JADES field at the same redshift slice \cite{2025MNRAS.539.1796L}. In the inner $21\arcsec$ ($\approx125$~kpc) region (coinciding with the detected X-ray emission), the overdensity increases to $\delta_{gal} = 4.5$. This corresponds to a $4.2\sigma$ overdensity detection (for details see Methods Section~1). At $z\approx5.7$, this overdensity is exceptionally rare.  Above $z>5$, the expected amplitude of galaxy overdensities falls off sharply due to the declining galaxy number density, increasing photometric incompleteness, and the fact that large‐scale structure has had less time to collapse. While the large galaxy overdensity associated with JADES-ID1 indicates a nascent proto-cluster, X-ray data can confirm whether the virialization has begun. Indeed, detecting hot ICM, best accomplished through sensitive X-ray imaging, can trace the gravitational potential of the proto-cluster and allows a more precise determination of its centroid, luminosity, and total mass.

\begin{figure}[!t]  
\centering
  \includegraphics[width=1.00\textwidth]{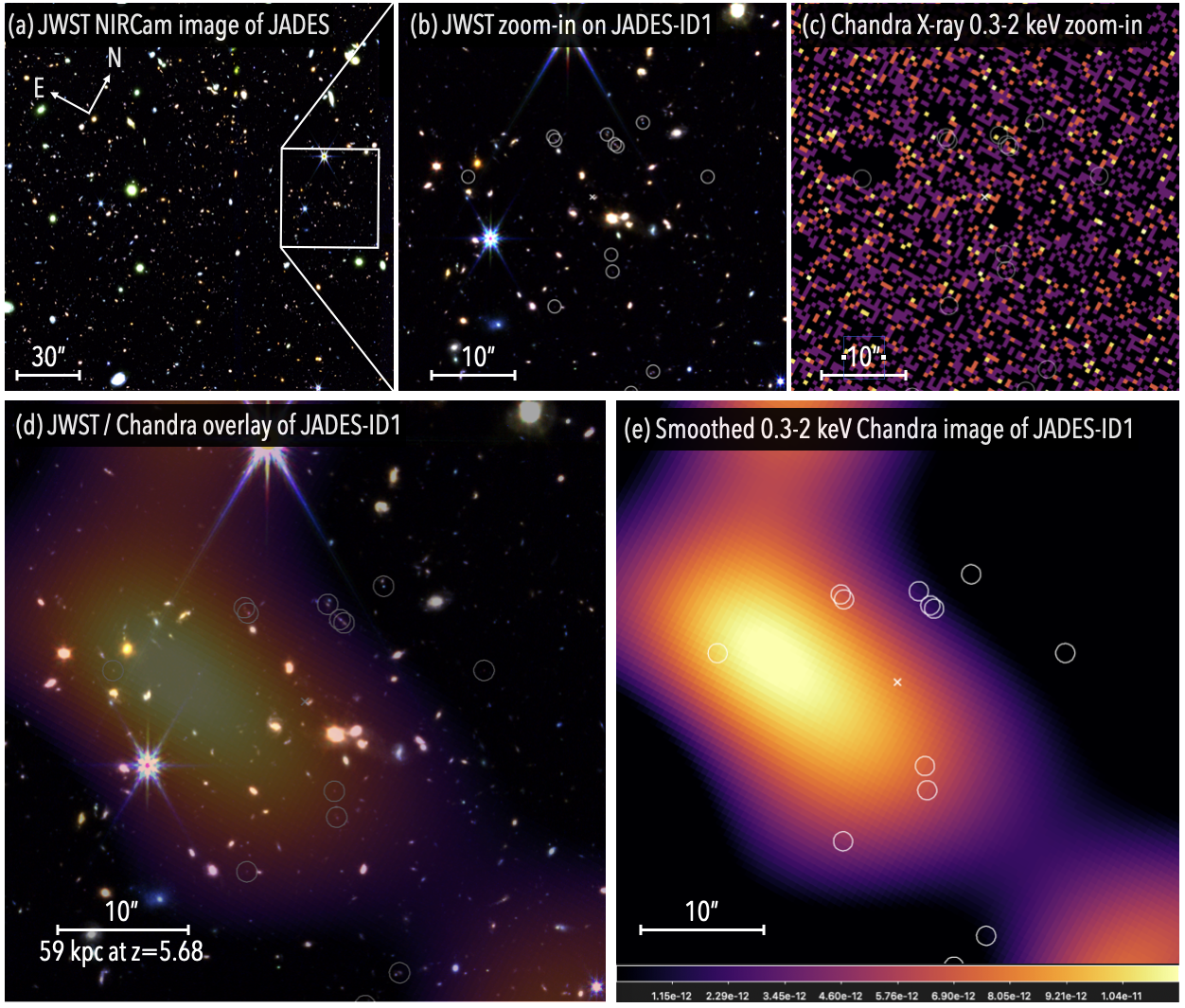}
 \caption{\textbf{Multi-wavelength view of the JADES-ID1 protocluster and its surroundings.} Panel (a) shows a composite \textit{JWST} image of the JADES field with a $45\arcsec\times45\arcsec$ ($265\times265$~kpc) box marking JADES-ID1. Panel (b) zooms in on this region, while panel (c) displays the exposure-corrected $0.3-2$~keV band \textit{Chandra} image of the same region. Panel (d) presents a \textit{JWST}/\textit{Chandra} overlay of JADES-ID1, where the \textit{Chandra} image has been processed by filling point sources, subtracting the background, and applying Gaussian smoothing with a kernel size of 15 pixels. Panel (e) shows the same processed \textit{Chandra} image, with the locations of likely cluster member galaxies highlighted. The \textit{Chandra} image reveals extended X-ray emission that is spatially coincident with the galaxy overdensity identified by \textit{JWST}.}
 \label{fig:image}
\end{figure}

To probe the X-ray-emitting properties of JADES-ID1, we used deep \textit{Chandra} Advanced CCD Imaging Spectrometer (ACIS-I) X-ray observations of the Chandra Deep Field South (CDFS). The CDFS overlaps with the JADES field and is the site of the deepest X-ray observation ever conducted \cite{2002ApJS..139..369G,2017ApJS..228....2L}. The data were analyzed with CIAO tools \cite{2006SPIE.6270E..1VF}. First, the 99 individual observations (Table~\ref{tab:obsids}) were reprocessed with the \texttt{chandra\_repro} tool. These observations were then merged into a single event file, which had a total exposure time of $6.55$~Ms. Next, we generated images and their corresponding exposure maps in the $0.5-2$~keV and $3-7$~keV energy ranges. Bright X-ray point sources were excluded from the analysis. A complete description of the data analysis is presented in Methods Section~2.

Figure~\ref{fig:image} presents a multi-wavelength view of the JADES-ID1 proto-cluster and its surroundings. While the \textit{JWST} data reveal a galaxy overdensity at $z\sim5.7$, an initial inspection of the exposure-corrected $0.3-2$~keV \textit{Chandra} image does not show bright X-ray emission co-spatial with JADES-ID1. It also indicates that none of the high-redshift galaxies are detected as individual X-ray point sources, and a stack of the galaxy positions likewise shows no statistically significant signal (Methods Section~4). To better assess the presence of faint diffuse emission expected from a $z\sim5.7$ proto-cluster, we carry out an in-depth imaging analysis. To this end, we first filled the location of the excluded point sources using the \texttt{dmfilth} tool using the ``global'' method. Next, we subtracted the background components using an annulus with $40\arcsec-110\arcsec$ radii around JADES-ID1 (Methods Section~2). Finally, we applied Gaussian smoothing to the image with a kernel size of $15$ pixels. 

The resulting image, shown in Figure~\ref{fig:image}{\it e}, indicates that large-scale diffuse X-ray emission is present near the \textit{JWST}-derived centroid. We find that the X-ray centroid (RA=3:32:31.75, Dec=-27:46:51.5) is offset by $\sim8\arcsec$ ($\sim47$~kpc) from the galaxy overdensity peak. Such separations are commonly observed in dynamically young or merging systems. Systematic studies of disturbed clusters find mean X‐ray/optical offsets of $\sim75$~kpc, with offsets reaching reaching and exceeding $\sim100$~kpc in the most unrelaxed cases \cite{2012MNRAS.420.2120M,2016MNRAS.457.4515R,2023A&A...671A..57S}. Even high-redshift systems, which are still in the early stages of collapse or undergoing mergers, exhibit comparable offsets \cite{2016ApJ...817..122B,2023Natur.615..809D}. Furthermore, due the broad Gaussian smoothing kernel (15 \textit{Chandra} pixels or $7.4\arcsec$) applied to enhance faint emission, our X-ray centroid is uncertain at the few arcsecond level. The $0.3-2$~keV \textit{Chandra} image nonetheless clearly shows extended emission spatially coincident with the galaxy overdensity, suggesting the presence of a large-scale hot ICM. Motivated by the detection of extended emission, we quantify this emission through a multi-pronged approach: (i) we establish its extended nature by constructing a surface brightness profile, (ii) we probe its presence in different X-ray energy ranges, and (iii) we examine its spectral properties. We note that the unsmoothed, point source‐masked image was  used for the further quantitative analysis.

\begin{figure}[!t]  
\centering
  \includegraphics[width=0.9\textwidth]{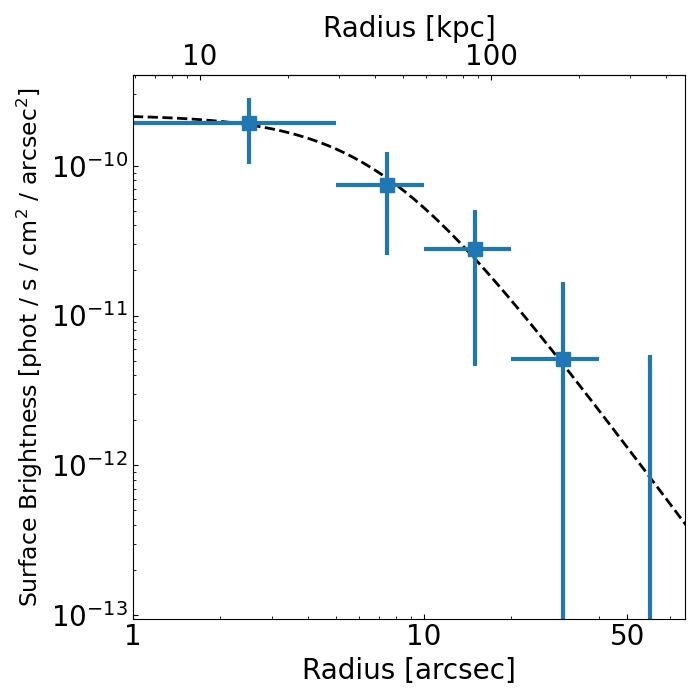}
\caption{\textbf{X-ray surface brightness profile of the JADES-ID1 proto-cluster.} The $0.3-2$~keV band profile was extracted in concentric annuli centered on the X-ray peak. The solid line shows the best‐fit $\beta$-model. The profile is corrected for exposure variations using the exposure map. The background is subtracted based on the $40\arcsec-110\arcsec$ ($235-646$~kpc at $z=5.68$) annulus around the proto-cluster. The resulting background level is $8.03 \times 10^{-10} \ \mathrm{photons \ s^{-1} \ cm^{-2} \ arcsec^{-2}}$. Extended emission is detected out to $\approx21\arcsec$ ($\approx125$~kpc). The error bars represent statistical uncertainties derived using the Gehrels approximation \cite{1986ApJ...303..336G}.}
 \label{fig:sb_profile}
\end{figure}

To quantify the large-scale diffuse ICM emission associated with JADES-ID1, we constructed an exposure-corrected, azimuthally averaged surface brightness profile from the unsmoothed, point source‐masked $0.3–2$~keV band \textit{Chandra} image, centered on the X-ray centroid. The background was measured in an annulus with $40\arcsec-110\arcsec$ ($235-646$~kpc at $z=5.68$) radii. For the background subtraction we accounted separately for vignetted (Milky Way foreground plus unresolved cosmic X-ray background) and non-vignetted (particle background) components by generating two exposure maps: one includes mirror vignetting and detector effects, and the other only includes detector effects. We applied each map to its corresponding background components (see Methods Section~2 for details). We also verified that the resulting surface brightness profile is insensitive to the exact choice of background region. The background-subtracted surface brightness profile, presented in Figure~\ref{fig:sb_profile}, reveals extended X-ray emission within $\approx21\arcsec$ ($\approx125$~kpc) and a declining trend with radius from the center. Beyond this radius, the signal-to-noise ratio declines and the emission becomes consistent with the background. Fitting this surface brightness profile with a $\beta$-model with fixed $\beta=0.6$, yields a core radius of $r_{\rm{c}} = 7.\!\arcsec1 \pm 3.\!\arcsec9$ ($42\pm23$~kpc). The observed $0.3-2$~keV band emission is significantly more extended than the \textit{Chandra} point spread function ($90\%$ of the encircled is contained within $1\arcsec$), confirming the truly extended nature of this emission. 

Within a $21\arcsec$ aperture ($\approx125$~kpc) of the \textit{JWST}-derived centroid, where the signal-to-noise ratio peaks, we measure $142\pm45$ net counts and 1858 background counts, supporting the detection of an extended ICM. If this emission originates from plasma with a few keV temperature, we do not expect to detect X-ray emission in the $3-7$~keV band. Indeed, at $z\approx5.68$, the observed $3-7$~keV band corresponds to the $20-46.8$~keV band in the rest-frame of the cluster, where such thermal emission is negligible. In agreement with this, the $3-7$~keV band surface brightness profile does not show statistically significant X-ray emission in the same aperture. We detect only $51\pm55$ net counts and 2947 background counts, consistent with the interpretation that the detected $0.3-2$~keV band emission originates from a hot ICM. 

Because of the relatively low count rate and the limited signal-to-noise ratio, extracting and fitting an X-ray spectrum of JADES-ID1 is not feasible. Instead, we investigate the spectral properties of the diffuse X-ray emission by calculating an X-ray hardness ratio $\mathrm{HR} = S / H$, where $S$ and $H$ correspond to the counts in the  $0.5-1.2$~keV and $1.2-2$~keV bands, respectively (for details see Methods Section~3). Within the same $21\arcsec$ source region, we obtain $\rm{HR}=1.84^{+1.96}_{-1.84}$. Although the large uncertainties only allow a broad estimate, these values imply an ICM temperature of at least $\sim2.5$~keV, consistent with a hot proto-cluster.

\begin{figure}[!t]  
\centering
  \includegraphics[width=0.98\textwidth]{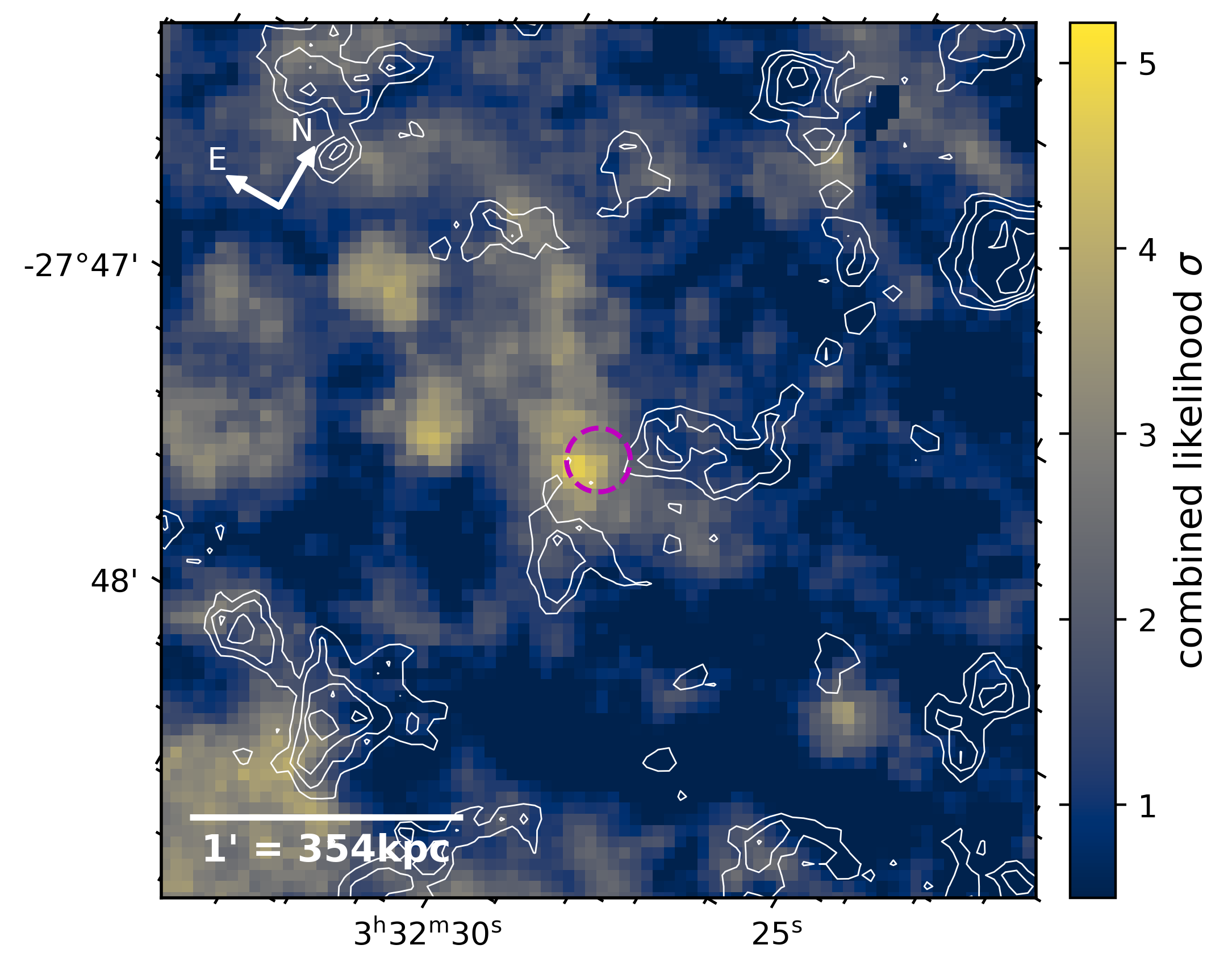}
 \caption{\textbf{Combined detection likelihood of JADES-ID1 and its surroundings.} White contours show the $3-7$~keV band emission. The map incorporates the probability of having a $0.3-2$~keV band detection, a $3-7$~keV band non-detection, and a radially declining surface brightness profile. The position of the proto-cluster is highlighted with a magenta circle with $7\arcsec$ radius.}
 \label{fig:significance_map}
\end{figure}

We next combine the key X-ray observables -- the $0.3-2$~keV band detection, the $3-7$~keV band non-detection, and the declining surface brightness profile -- into a combined likelihood map. This map quantifies the likelihood of random fluctuation mimicking all of these observational signatures simultaneously. For details on this map, we refer to Methods Section~5. In this map, shown in Figure~\ref{fig:significance_map}, JADES-ID1 stands out as the highest likelihood detection. Specifically, we find that the combined likelihood of detecting a statistical fluctuation with these parameters is $2.6\times10^{-7}$, which corresponds to a $5.0\sigma$ detection. While a few other regions also show elevated values on this map, those may trace structures at lower redshifts or they could represent statistical fluctuations. Thus, we conclude that the detected diffuse emission most likely originates from hot ICM associated with the JADES-ID1 proto-cluster. Finally, we combine the \textit{JWST}-based overdensity significance ($\approx4.2\sigma$) \cite{2025MNRAS.539.1796L} and \textit{Chandra} X-ray detection likelihood ($\approx5.0\sigma$) to establish the overall confidence level for JADES-ID1. We thus obtain a joint p-value of $3.4\times10^{-12}$, which corresponds to a $6.9\sigma$ detection.

Based on the detection of hot ICM associated with JADES-ID1, we derive the proto-cluster's X-ray luminosity and total mass. To this end, we use the net count rate within $21\arcsec$ radius, along with the \textit{Chandra} exposure maps to calculate the X-ray flux. We assume an optically-thin thermal plasma emission model with $kT=2$~keV temperature, $Z=0.3\ \rm{Z_{\odot}}$ metallicity \cite{2016ApJ...826..124M,2024MNRAS.527.1033B}, and a Galactic hydrogen column density of $N_{\rm H} = 6.7\times10^{19} \ \rm{cm^{-2}}$ \cite{2016A&A...594A.116H}. This yields an absorption-corrected $0.3-2$~keV band X-ray flux of $f_{\rm X}	= (4.6\pm{1.5}) \times10^{-16} \ \rm{erg \ s^{-1} \ cm^{-2}}$. Adopting the redshift of $z=5.68$ and applying a bolometric correction, we obtain a bolometric ($0.01-100$~keV) luminosity of $L_{\rm bol} = (1.5^{+0.5}_{-0.6}) \times10^{44} \ \rm{erg \ s^{-1}}$. To estimate the proto-cluster's total mass and ICM temperature, we assume self-similar evolution and apply the $L_{\rm bol} - M_{\rm 500}$ and $L_{\rm bol} - kT$ scaling relations obtained for ``all clusters'' in the sample of \cite{2020ApJ...892..102L}. These result in a gas temperature of $kT	= 2.7^{+0.5}_{-0.7}$~keV, which is consistent with the lower limit from the analysis of the X-ray hardness ratio. Finally, based on the luminosity we estimate a total cluster mass of $M_{500}= (1.8^{+0.6}_{-0.7}) \times 10^{13} \ \rm{M_{\odot}}$, which corresponds to $R_{\rm 500,c}\approx80$~kpc at $z=5.68$. We note that the uncertainties include the statistical errors on the flux measurements and the uncertainties on the scaling relation parameters. However, these estimates carry several caveats. First, the assumption of self-similar evolution may not fully capture deviations from the scaling relations at high redshift, where proto-clusters are still assembling and undergoing early-stage accretion. Second, if JADES-ID1 is experiencing mergers or shocks, its ICM temperature and  luminosity could be elevated \cite{2001ApJ...561..621R,2002MNRAS.329..675R,2007PhR...443....1M,2013MNRAS.429.2617O,2022ApJ...925...91S,2024arXiv240703142C}. However, such merger-driven boosts only add a modest bias in the scaling relations and do not qualitatively alter our conclusions \cite{2002MNRAS.329..675R,2006ApJ...639...64O,2011ApJ...729...45R}. Third, if the ICM temperature or metallicity differs from the assumed values, it could introduce a small change in the inferred parameters, although it would not alter our conclusions in any significant way. Finally, deviations from hydrostatic equilibrium, which are expected in dynamically young systems, could also impact the mass estimates, as non-thermal pressure support from turbulence or bulk flows may bias the derived masses low \cite{2012ApJ...758...74B,2016MNRAS.455.2936S}.

The \textit{JWST}-\textit{Chandra} detection of a proto-cluster and its hot ICM at $z=5.7$ provides insights into the formation of the first galaxy clusters and constrains the evolution of the hot ICM. JADES-ID1 is a rare example of a proto-cluster caught in the early phases of virial heating \cite{2014MNRAS.439.2146C,2024MNRAS.527.1033B}. Theoretical studies suggest that proto-cluster shock heating (virialization) typically begins at lower redshifts ($z\sim2-3$) \cite{2013ApJ...779..127C}. However, the presence of a hot ICM in JADES-ID1 demonstrates that, at least in some of the most massive proto-clusters, this process began merely one billion years after the Big Bang. While the detection of a hot ICM indicates the onset of gravitational collapse, the proto-cluster is unlikely to be fully virialized at this early stage. The signatures of ICM heating in JADES-ID1 may be linked to its exceptional richness. Indeed, with 66 potential members, it is by far the richest proto-cluster candidate in the sample of \cite{2025MNRAS.539.1796L}, suggesting that its deep gravitational potential could have accelerated both its collapse and the heating of its ICM. 

The detection of a massive proto-cluster with $M_{500}= (1.8^{+0.6}_{-0.7}) \times 10^{13} \ \rm{M_{\odot}}$ at such a high redshift is rather surprising. This detection provides an important data point for probing the abundance of large-scale structure systems in the early universe. To quantify how rare such systems are, we utilized the Tinker halo mass function and derived the expected number of massive halos within the $z=5-7$ redshift range \cite{2008ApJ...688..709T}. At these redshifts, the observable universe is predicted to host only a few dozen of systems with $M_{500} = 10^{13} \ \rm{M_{\odot}}$ and less than one with $M_{500} = 2\times10^{13} \ \rm{M_{\odot}}$ mass. Considering the small volume probed by the JADES/CDFS field, with the \textit{Chandra} pointings covering a footprint of $16'\times16'$ (or a comoving volume of $\approx1.3\times 10^6 \ \rm{Mpc^3}$), the probability of detecting a $10^{13} \ \rm{M_{\odot}}$ proto-cluster in this region is $\sim4\times10^{-5}$ under a $\Lambda$CDM cosmology \cite{2020A&A...641A...6P}. The probability drops even further to $\sim 2\times10^{-7}$ for a proto-cluster with a mass of $2\times10^{13} \ \rm{M_{\odot}}$. To further illustrate the improbability of $ \gtrsim 10^{13} \ \rm{M_{\odot}}$ halo, we note that the most massive halo in our survey volume is expected to be $\sim10^{12}\ \rm{M_{\odot}}$, about an order of magnitude below our X-ray-inferred mass. While these halo abundance probabilities assume a cosmic average, the matter-density variance on the scales of the JADES field only modestly alters the detection likelihood (Methods Section 1). Overall, the detection of the JADES-ID1 proto-cluster challenges our understanding of early-structure formation. This finding is analogous to recent \textit{JWST} discoveries revealing an overabundance of unexpectedly luminous galaxies at $z=9-12$ \cite{2023Natur.616..266L,2023NatAs...7..731B,2024MNRAS.531.2615C,2024arXiv240714973C,2025MNRAS.536.1018L}. However, we note that the presence of extended, shock-heated gas in a $\sim10^{13} \ \rm{M_{\odot}}$ halo is the direct consequence of gravitational collapse and virial shocks, offering a probe of rapid halo assembly. Taken together, these results provide further important clues that, in some regions, structure may have formed more rapidly than previously thought.

Continued multi-wavelength synergy is essential for mapping the first proto-clusters. While \textit{JWST} identifies high-redshift galaxy overdensities, next-generation X-ray missions could detect their extended ICM \cite{2019JATIS...5b1001G,2022arXiv221109827K,2025NatAs...9...36C}, while SZ experiments could reveal the thermal SZ imprint of the earliest proto-clusters \cite{2019BAAS...51g...6S,2023pcsf.conf..304C,2025A&A...694A.142M}.


\setcounter{figure}{0}
\newpage

\bmhead{Acknowledgements}
We thank Jake Bennett, Bill Forman, and Ralph Kraft for insightful discussions. \'A.B. and G.S. acknowledge support from the Smithsonian Institution and the Chandra Project through NASA contract NAS8-03060. This research has made use of data obtained from the Chandra Data Archive and software provided by the Chandra X-ray Center (CXC) in the application packages CIAO and Sherpa.
Q.L. and C.J.C. acknowledge support from the ERC Advanced Investigator Grant EPOCHS (788113). The \textit{JWST} part of this work is based on observations made with the NASA/ESA \textit{Hubble Space Telescope} (HST) and NASA/ESA/CSA \textit{JWST} obtained from the \texttt{Mikulski Archive for Space Telescopes} (\texttt{MAST}) at the \textit{Space Telescope Science Institute} (STScI), which is operated by the Association of Universities for Research in Astronomy, Inc., under NASA contract NAS 5-03127 for JWST, and NAS 5–26555 for HST. The observations used in this work are associated with JADES DR1 Release data of the GOODS-S field (PI: Eisenstein, N. Lützgendorf, ID:1180, 1210). 

\bmhead{Data Availability} This paper employs a list of Chandra datasets, obtained by the Chandra X-ray Observatory, contained in the Chandra Data Collection (CDC) 489~\href{https://doi.org/10.25574/cdc.489}{doi:10.25574/cdc.489}. 
The \textit{JWST} data of the JADES field is publicly available at \url{http://archive.stsci.edu}.

\bmhead{Code Availability} Data reduction and analysis employed standard, publicly available software packages: CIAO (Chandra Interactive Analysis of Observations) for processing and analyzing the X-ray data, Python with scientific libraries for handling the data and plotting, and SAOImage DS9 for image visualization.

\bmhead{Competing Interests} The authors declare that they have no competing financial interests.

\bmhead{Correspondence}  Correspondence and requests for materials should be addressed to \'A.B. (email: abogdan@cfa.harvard.edu).

\bmhead{Author contribution} \'A.B. analyzed the \textit{Chandra} observations, led the overall analysis, and drafted the manuscript. G.S. assisted with the \textit{Chandra} analysis, provided statistical tests and methods, contributed to the interpretation, and played a major role in writing the manuscript. Q.L. led the analysis of the \textit{JWST} data and contributed to the interpretation and text of the manuscript. C.J.C. contributed to the interpretation and text of the manuscript.

\newpage

\section*{Methods}

\section{\textit{JWST} characterization of the JADES-ID1 proto-cluster}
\label{sec:jwst}

Here we overview several key aspects of the \textit{JWST} analysis.  Specifically, we summarize the method used to define the JADES‑ID1 centroid, we quantify the statistical significance of its galaxy overdensity, and present tests that rule out the presence of any significant low‑redshift foreground structures.

To define the center of JADES-ID1, \cite{2025MNRAS.539.1796L} constructs two-dimensional galaxy overdensity maps in narrow redshift slices using the DETECTIFz algorithm \cite{2021MNRAS.506.2136S}, which employs Monte Carlo realizations of each galaxy's redshift probability distribution function. They then identify the slice with the highest overdensity peak, $z =5.68$ for JADES-ID1, and adopt the coordinates of that peak as the proto-cluster center.

Next, we quantify the rarity of the JWST‑identified overdensity around JADES‑ID1 by comparing it to field fluctuations.  Within a projected radius of $42\arcsec$ ($\approx250$~kpc) around the JADES‑ID1 centroid, the local galaxy overdensity is measured as $\delta_{\rm gal}=3.9$ relative to the mean density across the JADES field in the same redshift slice \cite{2025MNRAS.539.1796L}.  Focusing on the inner $21\arcsec$ ($\approx125$~kpc) region, which is coincident with our X‑ray aperture, the overdensity rises to $\delta_{\rm gal}=4.5$. This is comparable or exceeds those of previously confirmed proto‑clusters at similar redshifts \cite{2024ApJ...974...41H}. To assess statistical significance of this overdensity, we compared these values to the mean field density over $5.44<z<6.08$ within a spherical volume of radius $R_{200}=410$~kpc.  Accounting for a $\sim30\%$ cosmic variance, which is appropriate for the typical UV luminosities of candidate members \cite{2020MNRAS.499.2401T,2025MNRAS.539.1796L}, we performed $10^6$ Monte Carlo realizations.  The chance of obtaining the observed overdensity of the observed galaxies by random fluctuation is $1.4\times10^{-5}$, corresponding to a $\approx4.2\sigma$ detection.  We note that our field baseline includes cluster members, so both $\delta_{\rm gal}$ and its significance are slight underestimates relative to a purely field reference. This confirms that at $z\approx5.7$, such a strong overdensity is exceptionally rare.

While \cite{2025MNRAS.539.1796L} identify a clear overdensity at $z\approx5.68$, we further verify that no other significant structures exist at lower redshifts. To this end, we investigated the photometric galaxy catalogs based on \textit{JWST} and \textit{HST} observations in the JADES field. We measured the galaxy surface density within a $40\arcsec \times 40\arcsec$  box centered on the X‑ray emission peak, corresponding to the extent of the detected emission from the JADES‑ID1 proto‑cluster. We binned galaxies in redshift slices with width of $\Delta z =0.3$ in the redshift range of $z=0-6.6$ and derived the surface density in each bin. To correct for redshift‑dependent selection biases (most notably the increased completeness at lower redshifts), we carried out the same measurement in a large background region within the JADES footprint, while excluding the $40\arcsec \times 40\arcsec$  with the X-ray detection. The difference between the galaxy surface density at the position of the X-ray excess and the field average exhibits a single significant peak in the $z=5.25-6.23$ redshift bin. There are no comparable galaxy over-densities at any other redshift bins. This result demonstrates the absence of any significant foreground structure along the line of sight of the JADES-ID1 proto-cluster. 

The halo detection probabilities presented in the main body of the paper are based on the cosmic mean density field.  To estimate how local density fluctuations could bias these results, we estimate the variance of the matter density over a JADES‐sized volume.  We find $\sigma(R) \approx 0.059$, which corresponds to a $6\%$ typical fluctuation between different patches of this size. While a $6\%$ overdense field would allow structure to grow more rapidly, the chance of finding a $10^{13} \ \rm{M_{\odot}}$ halo would increase only from$\sim4\times10^{-5}$ to  $\sim3\times10^{-4}$.

\section{Chandra Data Analysis}
\label{sec:chandra_data}

To probe the presence and physical properties of the hot ICM associated with the proto-cluster JADES-ID1, we analyzed 99 \textit{Chandra} ACIS-I observations that cover the CDFS. The CDFS represents the deepest X-ray field ever observed. The list of analyzed \textit{Chandra} observations is given in Extended Data Table~\ref{tab:obsids}. The bulk of the data analysis was performed using standard CIAO tools, specifically, we used the latest version of CIAO (4.17) and the Calibration Database (CALDB 4.11.6). The main steps of the X-ray data analysis followed those outlined in our previous studies \cite{2022ApJ...927...34B,2024NatAs...8..126B}. Below, we outline the main steps of the X-ray analysis.

The first step of the analysis was to reprocess each individual observation using the \texttt{chandra\_repro} tool, thereby applying the latest calibration products. Next, we identified and filtered high-background periods from the observations using the \texttt{lc\_clean} routine by applying a $3\sigma$ threshold to remove fluctuations in the light curves. Since ACIS-I observations are not highly sensitive to solar flares, this step only reduced the exposure time by $\approx2\%$. The total cleaned exposure time of the dataset was $6.55$~Ms.

Because we analyze and combine a large set of observations, small differences exist in the alignment between the individual \textit{Chandra} observations. To account for this effect, we correct the absolute astrometry using the \texttt{wcs\_match} and \texttt{wcs\_update} tools. This step ensures that point sources are accurately aligned and exhibit a sharp point spread function, thereby  minimizing contamination of the extended emission. To perform the astrometry correction, we cross-matched the positions of X-ray point sources in individual observations with the coordinates of guide stars in the same field. Using the coordinates of the X-ray-optical source pairs, we applied frame transformations for each \textit{Chandra} observation, including transformations for rotation, scale, and translation. We set the deepest observation, ObsID 8594, as the reference coordinate system. For the subsequent analysis, we used these astrometry-corrected event files.

The next step in the analysis was to combine the individual X-ray observations. To this end, we used the \texttt{merge\_obs} tool, which co-added the data, resulting in a merged event file with $6.55$~Ms exposure time. This process also generated energy-filtered images in the $0.3-2$~keV (soft) and $3-7$~keV (hard) bands. In our analysis, we adopt the $0.3-2$~keV range as our soft band because it maximizes sensitivity to a few-keV thermal plasma at $z\approx5.7$. Since most observations were taken relatively early in the \textit{Chandra} mission, molecular contamination is minimal \cite{2018SPIE10699E..6BP,2022SPIE12181E..6XP}, allowing us to extend reliably down to 0.3~keV. For the hard band, we use the $3-7$~keV band, thereby avoiding $2-3$~keV energy range where the Au~L fluorescence line complex dominates \cite{2014A&A...566A..25B,2021A&A...655A.116S}. This hard band is sensitive to unresolved, AGN, very hot ICM at high redshift, or ICM emission from nearby clusters. We generated exposure maps for both energy ranges using this tool. These maps account for vignetting, molecular contamination, gaps between CCDs, and the filtering of bad pixels. To construct the exposure maps, we assumed an optically-thin thermal plasma model (\textsc{apec} in \textsc{XSpec}) with a Galactic column density of $N_{\rm H} = 6.8\times10^{19} \ \rm{cm^{-2}}$, a  $kT=2$~keV temperature, and $Z=0.3 \ \rm{Z_{\odot}}$ solar abundance. The count images were then divided by the exposure maps, producing the exposure-corrected images used in the analysis.

\begin{figure}[!t]  
\centering
  \includegraphics[width=0.98\textwidth]{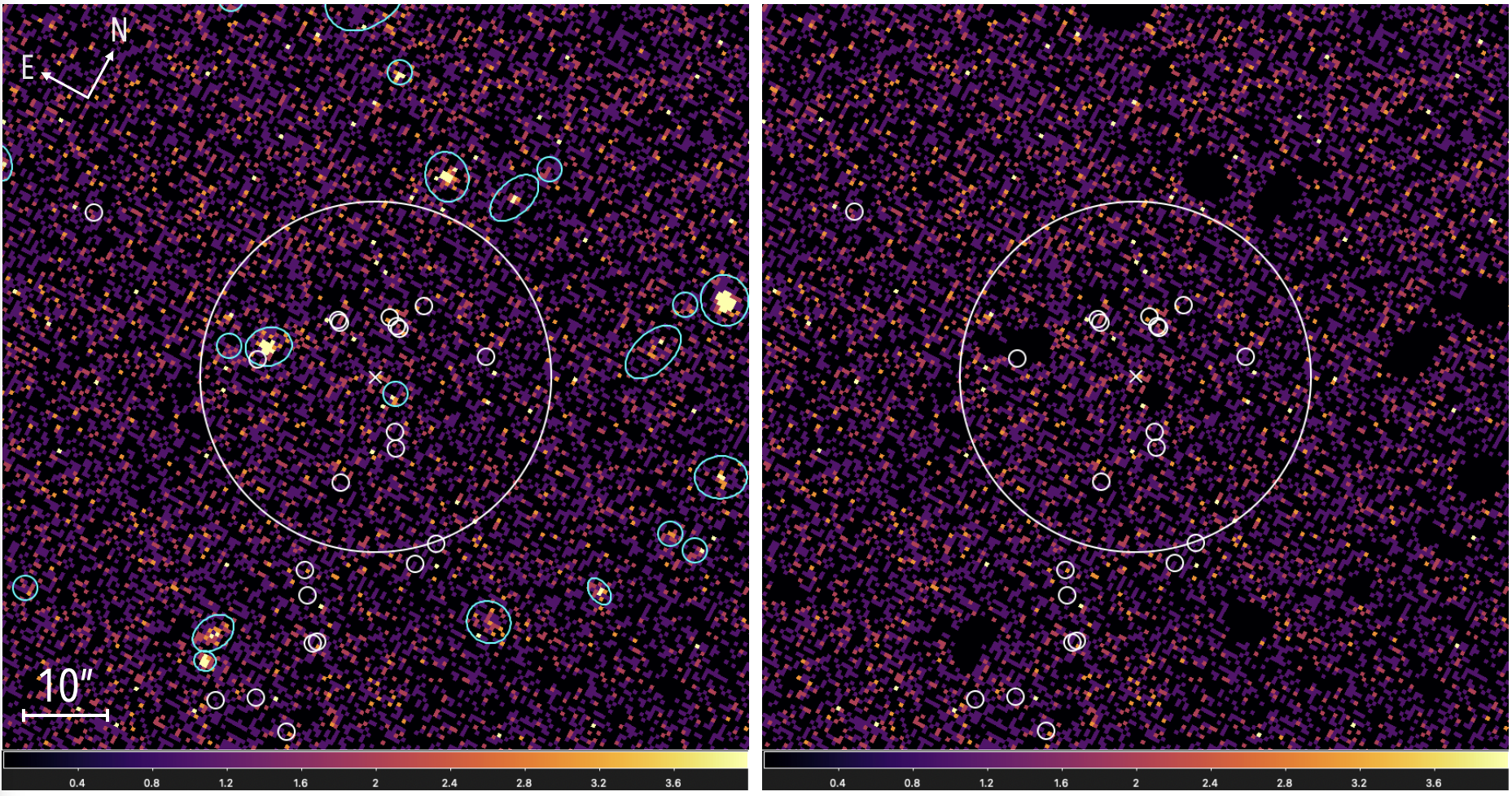}
\captionsetup{name=Extended Data Figure}
\caption{\textbf{\textit{Chandra} $\mathbf{0.3-2}$~keV band images of the JADES-ID1 field.} \textit{Left:} Merged \textit{Chandra} image with detected X-ray point sources marked by cyan regions and candidate $z\approx5.7$ member galaxies indicated by white circles. The large white circle with $21\arcsec$ radius denotes the aperture used for ICM analysis. \textit{Right:} The same \textit{Chandra} image after all X-ray point sources have been excised. }
 \label{fig:regions}
\end{figure}

To study the extended emission associated with JADES-ID1, we must ensure that bright X-ray point sources do not contaminate the extended emission from the proto-cluster. Therefore, we searched for resolved X-ray point sources (mostly originating from the cosmic X-ray background) using the \texttt{wavdetect} tool. We searched for sources using the wavelet scales of $1.0$, $1.414$, $2.0$, $2.828$, $4.0$, $5.657$, and $8.0$, which allows the detection of sources on a wide range of spatial scales and a significance threshold of $10^{-6}$. Additionally, we set the \texttt{ellsigma} parameter to 5. To carry out a comprehensive masking of all point sources, we cross‐referenced our source list with the CDFS point source list presented in \cite{2017ApJS..228....2L}. We found that some of their faintest sources, detected using a lower significance threshold, fell below our initial detection criteria, so we added these to our point source catalog. We then visually inspected the source regions and, for especially bright sources, enlarged the exclusion regions to fully encompass their PSF wings. We note that within the inner $21\arcsec$ of JADES-ID1, only one moderately bright and two faint point sources are detected (Extended Data Figure~\ref{fig:regions}). The detected point source regions were excluded from the merged images. Given the sharp \textit{Chandra} PSF, our large exclusion regions account for $\gtrsim96\%$ of the counts from each point source, implying that any residual counts have a negligible impact on the detected diffuse emission. Moreover, because spillover counts from point sources contribute similarly to both the source and background regions, it does not bias our measurements in any significant way. 

Since the X-ray emission from JADES-ID1 is relatively faint, precisely accounting for the background components is essential. The \textit{Chandra} background comprises multiple components: the vignetted sky emission (Milky Way foreground and unresolved cosmic X-ray background) and non-vignetted particle background. To separate these components, we generated two sets of exposure maps. One of them is a ``full'' map including mirror vignetting and all ACIS detector corrections, while the other is  a ``detector-only'' map that does not include the vignetting term. We then split the total background accordingly and applied the full map to the sky component and the detector-only map to the particle component. In the deep CDFS field, \textit{Chandra} resolves $\approx90\%$ of the cosmic X-ray background, so we find that particle background accounts for  $\sim77\%$ of the $0.3-2$~keV band and $\sim95\%$ of the $3-7$~keV band background.  

To verify that the X‑ray detection associated with JADES‑ID1 is not driven by a small subset of observations, we split the CDFS Chandra data into two groups in two complementary ways. First, we divided the observations chronologically into two sets with approximately equal total exposure times. Due to the buildup of the molecular contaminant on the ACIS optical blocking filter (which absorbs soft X-rays \cite{2018SPIE10699E..6BP,2022SPIE12181E..6XP}) and the soft, redshifted spectrum of the JADES‑ID1 proto‑cluster, the earlier data set contains $\approx65\%$ of the net counts, while the later one includes $\approx35\%$. Second, we randomly split the observations into two groups once again with approximately equal exposure times. In this case, since both subsets have the same average molecular contamination level, the net counts split roughly evenly between them. These tests demonstrate that the detected X‑ray signal is not driven by a handful of observations or by temporal variation but arises uniformly across the \textit{Chandra} dataset. This supports the conclusion that the X‑ray emission associated with JADES‑ID1 originates from a genuine, extended intracluster medium.

\renewcommand{\arraystretch}{1}
\begin{longtable}{ccc|ccc}
\captionsetup{name=Extended Data Table}
\caption{List of the analyzed \textit{Chandra} observations} 
\label{tab:obsids} \\
\hline
\textbf{ObsID} & \textbf{Exp (ks)} & \textbf{Date} & \textbf{ObsID} & \textbf{Exp (ks)} & \textbf{Date} \\
\hline
\endfirsthead

\hline
\textbf{ObsID} & \textbf{Exp (ks)} & \textbf{Date} & \textbf{ObsID} & \textbf{Exp (ks)} & \textbf{Date} \\
\hline
\endhead
\hline
441   & 55.97  & 2000-05-27 & 12233 & 35.57  & 2010-07-16  \\ 
582   & 130.58 & 2000-06-03 & 12234 & 49.15  & 2010-07-22  \\ 
1672  & 95.13  & 2000-12-16 & 16175 & 53.09  & 2014-10-03  \\ 
2239  & 130.84 & 2000-12-23 & 16176 & 24.68  & 2014-10-02  \\ 
2312  & 123.69 & 2000-12-13 & 16177 & 126.41 & 2014-10-08  \\ 
2313  & 130.40 & 2000-12-21 & 16178 & 73.98  & 2014-10-07  \\ 
2405  & 59.63  & 2000-12-11 & 16179 & 29.98  & 2014-12-31  \\ 
2406  & 29.68  & 2000-12-10 & 16180 & 49.44  & 2014-06-22  \\ 
2409  & 68.98  & 2000-12-19 & 16181 & 70.28  & 2014-10-31  \\ 
8591  & 45.43  & 2007-09-20 & 16182 & 75.06  & 2014-10-28  \\ 
8592  & 86.64  & 2007-10-22 & 16183 & 98.78  & 2014-06-09  \\ 
8593  & 49.49  & 2007-10-06 & 16184 & 55.38  & 2014-10-26  \\ 
8594  & 141.40 & 2007-11-01 & 16185 & 48.43  & 2016-03-24  \\ 
8595  & 115.42 & 2007-10-19 & 16186 & 29.40  & 2014-11-02  \\ 
8596  & 115.12 & 2007-10-24 & 16187 & 28.07  & 2014-11-03  \\ 
8597  & 59.28  & 2007-10-17 & 16188 & 103.70 & 2014-11-13  \\ 
9575  & 108.69 & 2007-10-27 & 16189 & 90.46  & 2014-11-29  \\ 
9578  & 38.57  & 2007-10-30 & 16190 & 116.74 & 2014-11-22  \\ 
9593  & 46.43  & 2007-09-22 & 16191 & 83.74  & 2015-05-25  \\ 
9596  & 111.89 & 2007-11-04 & 16450 & 81.04  & 2014-11-18  \\ 
9718  & 49.38  & 2007-10-03 & 16451 & 112.60 & 2015-03-24  \\ 
12043 & 129.58 & 2010-03-18 & 16452 & 27.71  & 2015-12-12  \\ 
12044 & 99.53  & 2010-03-23 & 16453 & 70.28  & 2015-03-21  \\ 
12045 & 99.72  & 2010-03-28 & 16454 & 47.13  & 2014-10-01  \\ 
12046 & 78.02  & 2010-04-08 & 16455 & 89.60  & 2015-10-27  \\ 
12047 & 10.14  & 2010-04-12 & 16456 & 47.46  & 2014-07-29  \\ 
12048 & 138.10 & 2010-05-23 & 16457 & 45.98  & 2014-08-05  \\ 
12049 & 86.94  & 2010-05-28 & 16458 & 96.38  & 2015-10-30  \\ 
12050 & 29.66  & 2010-06-03 & 16459 & 72.12  & 2015-06-20  \\ 
12051 & 57.29  & 2010-06-10 & 16460 & 21.44  & 2015-06-16  \\ 
12052 & 110.41 & 2010-06-15 & 16461 & 129.37 & 2015-05-19  \\ 
12053 & 68.11  & 2010-07-05 & 16462 & 143.91 & 2014-10-14  \\ 
12054 & 61.00  & 2010-07-09 & 16463 & 53.22  & 2014-09-23  \\ 
12055 & 80.68  & 2010-05-15 & 16620 & 33.71  & 2014-10-10  \\ 
12123 & 24.79  & 2010-03-21 & 16641 & 46.53  & 2014-07-31  \\ 
12128 & 22.80  & 2010-03-27 & 16644 & 44.01  & 2014-08-06  \\ 
12129 & 77.14  & 2010-04-03 & 17416 & 52.40  & 2014-09-28  \\ 
12135 & 62.53  & 2010-04-06 & 17417 & 12.67  & 2014-09-25  \\ 
12137 & 92.78  & 2010-04-16 & 17535 & 121.92 & 2014-10-17  \\ 
12138 & 38.53  & 2010-04-18 & 17546 & 19.82  & 2014-11-02  \\ 
12213 & 61.29  & 2010-05-17 & 17552 & 49.91  & 2015-10-10  \\ 
12218 & 87.98  & 2010-06-11 & 17556 & 46.87  & 2014-12-09  \\ 
12219 & 33.66  & 2010-06-06 & 17573 & 39.57  & 2015-01-04  \\ 
12220 & 48.13  & 2010-06-18 & 17633 & 35.52  & 2015-03-16  \\ 
12222 & 30.64  & 2010-06-05 & 17634 & 9.27   & 2015-03-19  \\ 
12223 & 100.71 & 2010-06-13 & 17677 & 108.73 & 2015-11-15  \\ 
12227 & 54.32  & 2010-07-14 & 18709 & 16.86  & 2015-11-22  \\ 
12230 & 33.81  & 2010-07-11 & 18719 & 34.53  & 2015-12-10  \\ 
12231 & 24.72  & 2010-07-12 & 18730 & 29.68  & 2016-02-02  \\ 
12232 & 32.89  & 2010-07-18 &  &  &  \\ \hline
\end{longtable}

\section{X-ray hardness ratios}
\label{sec:hardness}

Due to the relatively low number of net X-ray counts and the limited signal-to-noise ratios associated with the JADES-ID1 proto-cluster, it is not feasible to fit a full X-ray spectrum. Instead, we measure a hardness ratio (HR) to gain insight into the spectral properties of the emission. We define $\rm{HR} = S/H$, where $S$ and $H$ are the net counts obtained in the $0.3-1.2$~keV and $1.2-2$~keV bands, respectively. 

We estimate HR and its associated uncertainties using the Bayesian Estimation of Hardness Ratios (\texttt{BEHR}) code \cite{2006ApJ...652..610P}, which employs a Bayesian framework designed for low-count data. For JADES-ID1, this yields $\rm{HR}=1.84^{+1.96}_{-1.84}$. We note that \texttt{BEHR} evaluates the uncertainties through $10^6$ draws of Gibbs sampling that take into account the background counts. 

To explore how the measured HR constrains the ICM temperature, we generated synthetic HR values using \textsc{apec} models and using the exposure averaged response files. When deriving the HR values, we covered a temperature range of $kT=1-10$~keV and a metallicity range of $Z=0-1 \ \rm{Z_{\odot}}$. Within $kT\approx1-6$~keV range, the HR is only weakly sensitive to metallicity. On the other hand, the HR is more sensitive to temperature: hotter plasmas yield stronger emission above $1.2$ keV (rest-frame $8$~keV) and hence produce lower HR values. The upper limit of the measured $\rm{HR}\sim3.8$ suggests that the ICM temperature is at least $kT\sim2.5$~keV, although a higher temperature is also possible. This is consistent with our X-ray luminosity-based estimates, but the large HR uncertainties do not allow tighter constraints on either temperature or metallicity. Hotter ICM temperatures may imply additional heating from the ongoing mergers and other non-thermal processes (such as turbulent motions), both of which are expected in dynamically assembling clusters.

\section{Excluding Alternative Origins of the X-ray Emission}
\label{sec:scenarios}

Although the diffuse X-ray properties of JADES-ID1 consistently demonstrate that it originates from hot ICM, we have tested, and ruled out, potential non-thermal or point source origin of this emission. 

First, we cross-matched all resolved X-ray point sources with the positions of the candidate JADES-ID1 members within the source aperture (Figure~\ref{fig:regions}). None of these high-redshift galaxies coincides with a resolved \textit{Chandra} point source, implying that these galaxies do not host luminous AGN. Next, to probe whether a population of fainter, individually undetected AGN is responsible for the diffuse emission, we stacked the \textit{Chandra} counts at the location of each $z\approx5.7$ galaxy within the source aperture. To this end, we extract the source counts using a $1\arcsec$ radius aperture and a local $3\arcsec-6\arcsec$ annulus for background. This source region encircles $\sim90\%$ of the source counts at the position of JADES-ID1. This analysis results in stacked net counts of $-0.2\pm8.2$ in the $0.3-2$~keV band and $-10.9\pm9.2$ counts in the $3-7$~keV band. We note that the non-detection of high-redshift AGN is consistent with previous analyses showing that the vast majority of high-redshift galaxies are X-ray faint \cite{2024ApJ...969L..18A,2025MNRAS.538.1921M,2025arXiv250509669S}, with only rare exceptions \cite{2024NatAs...8..126B,2024ApJ...965L..21K}. Thus, the absence of detection both individually and in stack firmly rules out any significant contribution from either resolved or unresolved AGN.

In a recent study, \cite{2021ApJ...913..110C} suggested that inverse-Compton (IC) scattering of cosmic microwave background photons off radio galaxy electrons can effectively mimic the X-ray appearance of a high-redshift galaxy cluster. To test this scenario for JADES-ID1, we examined the MeerKAT $1.28$~GHz map of the CDFS field \cite{2025MNRAS.536.2187H}. 

At the location of the X-ray peak we find no significant radio emission above the noise level. Adopting a typical intracluster magnetic field \cite{2010A&A...513A..30B,2021NatAs...5..268D} of $4 \rm{\mu G}$ [$2 \rm{\mu G}$], IC scattering sufficient to reproduce the observed X-ray flux would imply a 1.4~GHz radio surface brightness that exceeds the MeerKAT detection limit by more than an order of magnitude ($>12 \sigma$ [$\sim 4\sigma$] per MeerKAT beam), which is not observed. 
We do, however, identify a faint radio point source $11.5\arcsec$ offset from the X-ray peak with a flux density of $13.7\pm1.4 \ \rm{\mu Jy}$. This radio source is coincident with a galaxy at $z=1.173$ (CANDELS J033230.91-274649.5) \cite{2023ApJ...945..155M} and not with any of the $z\approx5.7$ potential cluster members. 
At this position, a faint X-ray point source is detected, which has been excised from the analysis of the diffuse emission. We do not expect significant X-ray IC emission from the radio lobes of this galaxy. 
Even if we (incorrectly) attribute the entire radio flux at $z=5.7$, we expect only $\sim5$ [$\sim 15$] IC X-ray counts in either the soft and the hard band, far below the observed X-ray signal. Thus, IC emission cannot account for the observed soft X-ray emission. 

Taken together, the absence of any X-ray point source coinciding with a JADES-ID1 cluster member, the non-detection from the stacking analysis, and the inconsistency between predicted IC flux emission and the observed flux as well, all demonstrate that neither AGN nor IC scattering can explain the extended soft X-ray emission. Therefore, the only viable explanation remains thermal bremsstrahlung from a few-keV ICM at $z\approx5.7$ associated with the JADES-ID1 proto-cluster.

\section{Detection significance map}
\label{sec:significance_map}

The detection of extended X-ray emission in the $0.3-2$~keV band, along with the declining surface brightness profile, and the $3-7$~keV band non-detection all provide strong evidence for the presence of hot ICM associated with JADES-ID1.

Combining all this information, we construct a likelihood map to quantify the probability of hot intracluster gas associated with JADES-ID1 (Figure~\ref{fig:significance_map}). 
We define the combined likelihood for a high redshift cluster as follows, 
\begin{equation}
    \mathcal{L} = \mathcal{S}_{\rm s}(r<6\arcsec) \times \left[ \mathcal{S}_{\rm s} (6<r<20\arcsec) \times \mathcal{S}_{\rm s} (0<r<6\arcsec) \right]^{0.5} \times \mathcal{C}_{\rm h} (<6\arcsec),
\end{equation}
where $\mathcal{S}$ is the Poisson survival function and $\mathcal{C}$ is the cumulative distribution function. The subscripts, $s$ and $h$ denotes the soft and hard band, respectively. 
The first part, $\mathcal{S}_{\rm s}(r<6\arcsec)$, defines the probability of detecting counts within 6$\arcsec$ above the background level ($50\arcsec-75\arcsec$). The second part, $\mathcal{S}_{\rm s} (6\arcsec<r<20\arcsec)$ describes the detection of counts between $6\arcsec-20\arcsec$, but the background is assumed to be the $20\arcsec-30\arcsec$ region. The third part, $\mathcal{S}_{\rm s} (0\arcsec<r<6\arcsec)$, analogous to the second part, describes the excess counts in the inner 6$\arcsec$, but used the $6\arcsec-20\arcsec$ region as background. Together, the second and third part will detect a rising profile toward the center. The last part, $\mathcal{C}_{\rm h} (<6\arcsec)$ brings in the probability of having hard band counts within 6$\arcsec$ consistent with the background ($50\arcsec-75\arcsec$). We note that for the $\beta$-model shown in Figure~\ref{fig:sb_profile}, the choice of $0\arcsec-6\arcsec$ and $6\arcsec-20\arcsec$ bins maximizes the signal-to-noise ratio, and makes it also equal in the two bins.

For plotting purposes we show the combined likelihood converted into a significance, using the inverse survival function. This map reveals that the combined likelihood of a statistical fluctuation with these parameters is $2.6\times10^{-7}$, which corresponds to a $5.0\sigma$ detection. This indicates that the diffuse X-ray emission most likely originates from hot ICM associated with JADES-ID1. In Figure~\ref{fig:significance_map}, JADES-ID1 is the most significant detection. Specifically, this is the only region that satisfies all of our X-ray criteria (a soft band detection, a hard band non-detection, and a declining surface brightness profile) and is co-spatial with the independently identified \textit{JWST} galaxy overdensity. All other apparent ``hot spots'' on this map either lie outside the $z\approx5.7$ redshift slice or fail one or more of the X-ray criteria required to identify a high-redshift galaxy cluster, indicating that these regions are statistical fluctuations or correspond to lower-redshift structures tracing the large-scale structure of the universe.





\bibliographystyle{naturemag}
\bibliography{paper1.bib} 


\end{document}